\documentclass{ws-procs975x65}

\begin{document}

\title{Limits on the dark matter particle mass from black hole growth in galaxies} 

\author{Faustin Munyaneza\footnote{Humboldt Fellow.}}

\address{Max-Planck-Institut f\"ur Radioastronomie,\\
69 Auf Dem H\"ugel\\
D-53121 Bonn, Germany\\
\email{munyanez@mpifr-bonn.mpg.de}}

\begin{abstract}
I review the properties of degenerate fermion balls and 
investigate the dark matter distribution 
at galactic centers using NFW, Moore and isothermal 
density profiles. I show that 
dark matter becomes degenerate for particles masses 
of a few keV  at  distances less than a few 
parsec from the center of our galaxy. 
To explain the galactic center black hole of mass of $\sim 3.5 \times 10^{6}M_{\odot}$ and 
a supermassive black hole of $\sim 3 \times 10^{9}M_{\odot}$ at a redshift
of 6.41 in SDSS quasars, the mass of the fermion ball is assumed  to 
be between $3 \times 10^{3} M_{\odot}$ 
and $3.5 \times 10^{6}M_{\odot}$. 
This constrains  
  the mass of the dark matter particle
 between   $0.6 \ {\rm keV}$ 
 and $82~{\rm keV}$.
 The lower limit on the dark matter mass 
  is improved to about {\rm 6~keV} if exact solutions of Poisson's equation
  are used in the isothermal power law case. The constrained dark matter particle
  could be interpreted as a sterile neutrino.
\end{abstract}
\bodymatter

\section{Introduction}\label{intro}
  There is 
 mounting evidence that
 most galaxies harbor supermassive black holes (BHs) of masses 
 from $10^{6.5}$ to $10^{9.5} M_{\odot}$.
 The typical case is of the Galactic black hole of mass 
 $(3.1 \pm 0.9) 10^{6} M_{\odot}$~\cite{schodel03,ghez05}.
 It has also been established that
the mass of the central black hole 
is tightly correlated with the velocity dispersion $\sigma$ of its host bulge,
where it is found that $M_{BH} \sim  \sigma^{4-5}$~\cite{fm00}.
In spite of the vast and tantalizing work on black hole
physics, their genesis and evolution are not well understood.
Another outstanding problem in modern astroparticle physics 
is the particle nature of dark matter(DM).
Recently, there has been a renewed interest in sterile neutrinos\cite{dw94} as 
candidates for dark matter as they could explain the baryon asymmetry of the Universe\cite{asaka05}, the pulsar kicks\cite{kusenko04},
the early growth of black holes \cite{mb05,mb06,mbgc06} and the 
reionisation of the Universe\cite{bk06,jarek06}.
In this short communication, we review the properties fermion balls, i.e. self gravitating 
systems of sterile
neutrinos and study the implications of NFW\cite{nfw97}, Moore\cite{moore99} and 
isothermal density profiles being degenerate
near the galactic centers.

\section{Fermion balls and dark matter mass limits}

We will assume that the DM particles obey a Fermi Dirac
distribution function with a non vanishing chemical potential \cite{mb05}.
Assuming only gravitational interaction, it has been shown that these dark matter
 particles could
condensate at galactic centers  forming  degenerate fermi
 balls \cite{v94,tv98,mtv98,mtv99,bmv99,bmtv02,bmtv02b,mv02} with 
 a total mass $M_{F}$ that scales with their size $R_{F}$ as
  $M_{F} \sim m_{s}^{-8}R_{F}^{-3}$ where $m_{s}$ is the
 the sterile neutrino mass. Fermions have a maximum mass $M_{0V} \sim m^{3}_{pl} m_{s}^{-2}$ 
 where
 $m_{pl}=\left(\hbar c/G\right)^{1/2}$ is the Planck's mass.
 
It has been shown that the
assumption of a degenerate fermion ball of mass
 $M_{F}\sim 3\times 10^{6}M_{\odot}$ at the center of a
 dark matter halo of $3\times 10^{12}M_{\odot}$ with a density 
 scaling as $1/r^{2}$ in the outer edge of the halo
 constrains the fermion mass~\cite{mb05,mbgc06} to
 \begin{equation}
m_{s} \approx 12 \ {\rm keV/c}^{2} \left( \frac{\sigma}{156 \, {\rm km/s}}\right)^{3/4}
 \left( \frac{M_{bh}}{3 \ 10^{6}M_{\odot}}\right)^{-1/2} \left( \frac{2}{g_{s}}\right)^{1/4},
\label{eq:03}
 \end{equation}
where $\sigma$ is the velocity dispersion.
Using $\sigma = 100 \pm 20$ km/s \cite{fm00} for the velocity dispersion,
and the mass of the black hole of $M_{BH} \sim ( 2.2 - 4)10^{6} M_{\odot}$ \cite{schodel03},
 then
we obtain a lower limit on the mass of the DM particle 
 \begin{equation}
m_{s}  \stackrel {\textstyle >}{\sim} (6 - 12) \ {\rm keV/c^{2}} .
\label{eq:04}
\end{equation}
The above limits lie within the range of sterile neutrino masses obtained 
 from the study of the origin of the high velocities
up to 1000 km/s of pulsars \cite{kusenko04}. Moreover, similar limits were also obtained
from X-ray background studies \cite{mafe05}.
Recently, it has been shown that the
 decay of such a sterile
 neutrino could help
initiate star formation in the early Universe \cite{bk06,jarek06}.

The  black hole grows exponentially with time from
 Eddington limited baryonic matter accretion \cite{mb05}.
 However, unless the seed black hole has a high mass of
  $10^{3-4}M_{\odot}$, Eddington limited 
 baryonic matter accretion cannot grow the seed black holes 
 to $3 \times 10^{9}M_{\odot}$ black 
 holes in SDSS quasars\cite{willot03} at $z=6.41$. We have shown that the Pauli's degeneracy principle 
 helps 
 feed the black hole with dark matter \cite{mb05,mbgc06}.
  Stellar seed black 
 holes could be grown
 to $10^{3-4}M_{\odot}$
  in about $10^{7-8}$ years.
  For this model to work, the dark  matter 
  particle mass has been constrained to be in the order of 10 keV \cite{mb05}.
 A further growth
 to $10^{6.5-9.5}M_{\odot}$ is achieved through Eddington baryonic accretion.

 Another constraint on the dark matter particle mass is obtained by studying
the NFW, Moore and isothermal density profiles of $\rho \sim r^{-s}$ with $s$
 being the power law slope
and $s=1$ for NFW \cite{nfw97}, $s=1.5$ for 
Moore~\cite{moore99}
  and 
$s=2$ for the isothermal profiles. 
The mass enclosed within a radius $r$  scales as $M_{r} \sim r^{3-s}$ and 
the corresponding
rotational velocity scales as $v_{rot} \sim r^{1-s/2}$.
We then study the conditions under which these density profiles
 satisfy the Pauli
 degeneracy condition $m_{s} v_{rot} \sim \hbar \, n_{s}^{1/3}$, where $n_{s}$ is the sterile neutrino
 number density. It is found that for all the three density 
profiles\footnote{For NFW profile, the DM particle mass is
in the range 
of $ \ 0.6 \  {\rm keV} \stackrel {\textstyle <}{\sim} m_{s} \stackrel {\textstyle <}{\sim} 6 \ {\rm keV}$. 
A range
of  $ \ 1 \ {\rm keV} \stackrel {\textstyle <}{\sim} m_{s} \stackrel {\textstyle <}{\sim} 14 \ {\rm keV}$ 
is found
for Moore profile and finally we obtain a mass range of 
$2 \ {\rm keV} \stackrel {\textstyle <}{\sim} m_{s} \stackrel {\textstyle <}{\sim} 82 \ {\rm keV}$
for the isothermal density profile.}
 dark matter becomes
degenerate for particle masses between 0.6 keV and 82 keV \cite{mb06}.
The lower limit could be improved to about 6 keV if full solutions of Poisson's equation
are used \cite{mb05}.
 The mass of the degenerate core is
assumed to be between $3 \times 10^{3}M_{\odot}$ and $3.5 \times 10^{6}M_{\odot}$. The black holes
grow then by consuming the whole mass from degenerate cores 
and then  by Eddington baryonic matter accretion at late stages of
their evolution, i.e. at $t \stackrel {\textstyle >}{\sim} 10^{7-8}$ years.

\section{Conclusion}
Fermionic dark matter of mass of the order of a few keV could help boost 
the growth of supermassive black
holes at galactic centers. Moreover, NFW, Moore and isothermal density profiles
become degenerate for particles mass of a few keV at distances of 
a few parsec from the galactic
centers. The detection of an X-ray line at half the sterile neutrino by XMM-Newton and CHANDRA satellites
would be the smoking gun for the existence of a such sterile neutrino.

\section*{Acknowledgments}
It is a great pleasure to thank  Peter L. Biermann for encouragement
and support.  The author thanks the MG11 organisers for invitation
and the Humboldt foundation for financial support to attend the conference.


\begin{thebibliography}{00}

\bibitem{schodel03} R. Sch\"odel  et al., {\em ApJ} {\bf  596}, 1015 (2003)
\bibitem{ghez05} A. Ghez et al.,  {em ApJ} {\bf 620}, 744 (2005)
\bibitem{fm00} L. Ferrarese,  and D. Merritt,  {\em ApJ} {\bf 539}, L9 (2000)
\bibitem{dw94} S. Dodelson,  and L.M.  Widrow, {\em Phys. Rev. Lett.} {\bf 72}, 17 (1994)
\bibitem{asaka05} T. Asaka and M. Shaposhnikov, {\em  Phys. Rev. Lett.} {\bf  B620}, 17 (2005)
\bibitem{kusenko04} A. Kusenko, {\em Int. Journ. Mod. Phys.} {\bf D13}, 2065 (2004)
\bibitem{mb05} F. Munyaneza and P.L. Biermann, {\em Astr. \& Astrophy}, {\bf 436}, 805 (2005)
\bibitem{mb06} F. Munyaneza and P.L. Biermann, {\em  Astr. \& Astrophys.} {\bf  458}, L15-19 (2006)
\bibitem{mbgc06} F. Munyaneza and P. L. Biermann,  {\it Proceedings of the Galactic Center 
Workshop}, Bad Honnef, 18-22 April 2006,  Ed. R. Sch\"odel et al. {\it Journ. of Physics: Conference Series}, 
{\bf 54}, 456 (2006) 
\bibitem{bk06} P.L. Biermann and  A. Kusenko, {\em Phys. Rev.Lett.} {\bf  96}, 091301 (2006)
\bibitem{jarek06} J. Stasielak, A. Kusenko and  P. L.  Biermann,  {\em ApJ in press}, astro-ph/0606435 (2006)
\bibitem{nfw97} J.F. Navarro, C.S.  Frenk and S.D.M White, {\em ApJ} {\bf 490}, 493 (1997)
\bibitem{moore99} B. Moore, S. Ghigna, F.  Governato et al., {\em ApJ} {\bf 524}, L19 (1999) 
\bibitem{v94} R.D. Viollier, {\em Prog. Part. Nucl. Phys} {\bf  32}, 51 (1994)
\bibitem{tv98} D. Tsiklauri and R.D.  Viollier, {\em  ApJ} {\bf 500}, 591 (1998)
\bibitem{mtv98} F. Munyaneza, D.  Tsiklauri  and  Viollier, R.D, 
                 {\em ApJ} {\bf  509}, L105 (1998)
\bibitem{mtv99} F. Munyaneza, D. Tsiklauri and  R.D. Viollier,   {\em  ApJ} {\bf 526}, 744 (1999)
\bibitem{bmtv02} N. Bili\'c, F. Munyaneza, G. B. Tupper and R.D. Viollier, {\em Prog. Part. Nucl. Phys.} {\bf 48}, 291 (2002)
\bibitem{bmtv02b} N. Bili\'c, F. Munyaneza, G. B. Tupper and R.D. Viollier,
{\it Proceedings of International Conference DARK2002}, Cape Town, 19-22 Feb. 2002, p. 46-52,
ed. H.V.
Klapdor-Kleingrothaus, and R.D. Viollier, (Berlin Springer, 2002)
\bibitem{bmv99} Bili\'c, N., Munyaneza, F., Viollier, R. D. 1999, {\em Phys. Rev.} {\bf D. 59}, 024003 (1999)
\bibitem{mv02} F. Munyaneza and R.D.  Viollier, {\bf ApJ}  {\bf 564}, 274 (2002)
\bibitem{mafe05} M. Mappelli and  A. Ferrara, {\em  MNRAS} {\bf 364}, 2 (2005)
\bibitem{willot03} C.J. Willot, R.J. McLure and M. Jarvis, {\em ApJ} {\bf 587}, L15 (2003)
\end{thebibliography}
\end{document}